# Improvement and protection of niobium surface superconductivity by Atomic Layer Deposition and heat treatment.

T.Proslier[1,2], J. Zasadzinski[1], J.Moore[3], M.Pellin[2], J.Elam[4], L.Cooley[5], C.Antoine[6], J.Norem[3], K.E.Gray[2]

[1]Physic department, Illinois Institute of Technology, Chicago, IL 60616, USA
[2]Materials Science Division, Argonne National Laboratory, Argonne, IL 60493, USA
[3]High Energy Physic Division, Argonne National Laboratory, Argonne, IL 60493, USA
[4]Energy System Division, Argonne National Laboratory, Argonne, IL 60493, USA
[5]Technical Division, Fermi National Accelerator Laboratory, Batavia, IL 60510, USA
[6]Commissariat a l'énergie Atomique, Centre de Saclay, F-91191, Gif-Sur-Yvette, France

A method to treat the surface of Nb is described which potentially can improve the performance of superconducting RF cavities. We present tunneling and x-ray photoemission spectroscopy (XPS) measurements at the surface of cavity-grade niobium samples coated with a 3 nm alumina overlayer deposited by Atomic Layer Deposition (ALD). The coated samples baked in ultra high vacuum (UHV) at low temperature reveal at first degraded superconducting surface. However, at temperatures above 450C, the tunneling conductance curves show significant improvements of the superconducting density of states (DOS) compared with untreated surfaces.

For several decades, the performances of accelerating Niobium Superconducting Radio Frequency (SRF) cavities have been continually improved, to reach now a reproducible maximum accelerating field of ~30 MV/m and Quality factor Q above $10^{10}$. However unresolved mysteries remain (1) and prevent cavities from reaching the intrinsic Nb limits believed to be 55 MV/m. An SRF cavity is subject to an accelerating RF field $E(\omega)$ that induces an oscillating magnetic field $B(\omega)$ on the inner walls of the cavity. Penetration of the latter implies superconducting screening currents on a length scale $\lambda_{Nb}$, the niobium penetration depth (~45 nm at T=2 K). The dissipation mechanisms are confined to a region of depth $\lambda_{Nb}$, over which supercurrents interact strongly with the surface oxides. Up to recently (2), a detailed picture of these strong interactions and in particular the mechanisms by which oxygen influence the performance of niobium cavities was still missing.

The uncontrolled complex sets of oxides growing on the surface of air-exposed niobium leads to well known problematic effects such as weakened superconductivity, proximity effects or pair breaking phenomena. For niobium-based devices (3), sub-micron Josephson tunnel junctions or single electron transistors these problems have been cured by capping a cleaned niobium surface with alumina. However, at present there are no realistic ways to remove the oxides from the inner wall of niobium cavities and to deposit a uniformly thin protective layer on a complex shaped cavity. We propose a new approach to overcome these problems: the combination of the Atomic layer Deposition (ALD) technique with a subsequent annealing in UHV. We probe the occurrence of surface oxides with XPS and the near surface superconductivity by tunneling spectroscopy.

Tunneling spectroscopy and XPS measurements were done on air exposed cavity-grade niobium coupons capped with a protective alumina layer deposited by ALD (4). ALD is a self-limiting, sequential surface chemistry that has the unique ability to achieve uniform atomic scale deposition control on arbitrary complex-shaped substrates (5). The pinhole-free and dense structure of ALD made films make $Al_2O_3$ an ideal protective and diffusion barrier. Pieces of monocrystalline (110) and polycrystalline Nb were cut from larger sheets used to construct SRF cavities. These pieces were electropolished, cleaned with deionized purified water and dried in air in a manner similar to that done on cavities. They were later introduced into the ALD reactor and coated with 3 nm of $Al_2O_3$ at a temperature of 120°C under a flow of ultra pure nitrogen gas. A detailed description of the deposition process can be found in ref (6). The coated samples were baked in UHV at temperature ranging from 250 to 500 C and the XPS spectrum of the Nb 3d core level was measured simultaneously. After each thermal treatment, the samples were transferred in air to the point contact apparatus and the surface superconducting DOS was probed by tunneling spectroscopy at a temperature of 1.7 K as described elsewhere (7).

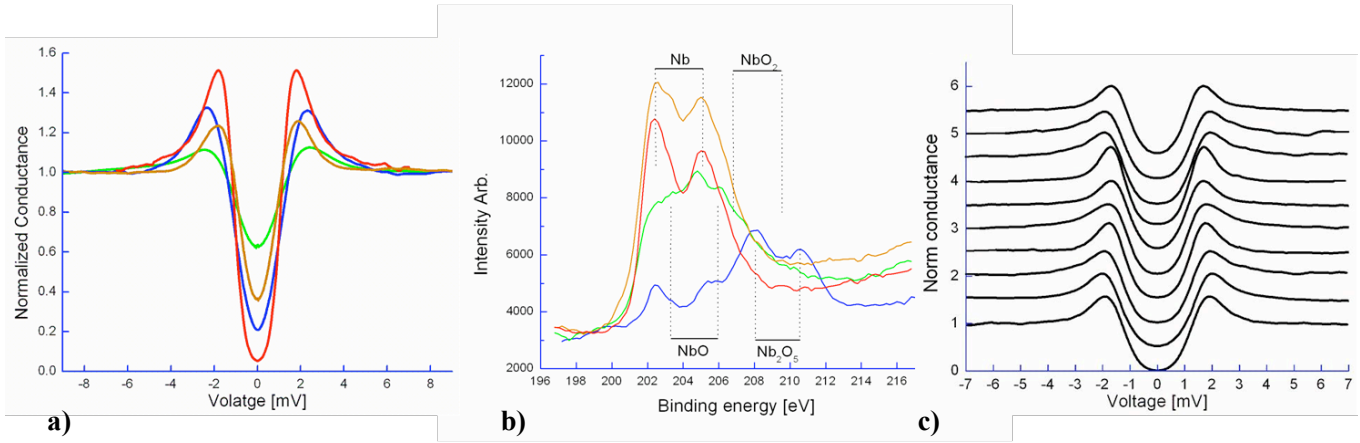

Fig.1: (Color online) **a)** conductance curves measured at 1.7K on a polycrystalline Nb sample coated with 3nm of $Al_2O_3$. **b)** Corresponding XPS spectrum of the Nb 3d core level. Each color corresponds to one temperature; 30 °C in blue, 220 °C in green, 380 °C in orange and 450 °C in red. **c)** Ten conductance curves measured on a coated Nb single crystal (110) baked at 500°C for 24h. The curves have been shifted for more clarity.

The evolution of the oxide's composition and the superconducting DOS at the surface of a polycrystalline sample as a function of the baking temperature is shown in Fig. 1: for annealing temperature up to 380°C, the tunneling spectroscopy (Fig. 1a) reveals degraded superconducting features as compared to an unbaked samples (in blue), with a zero bias conductance (ZBC) as high as 65% of the normal conductance (measured at V=12 meV). Importantly, for an anneal at 450°C for 24 h, the tunneling conductance becomes much sharper and the ZBC improves by a factor of 5 down to ~5%. In parallel, the XPS spectrum (Fig.1b), that probes ~3.5 nm depth on the surface, shows a reduction of $Nb_2O_5$ into sub oxides such as NbO and $NbO_2$ below 450°C (in green and orange), whereas at 450°C-24h the XPS unveils the presence of NbO only with sharper metallic $Nb^o$ peaks. The same reduction of the ZBC down to ~5% is found reproducibly on Nb single crystals (Fig.1c) baked at 500°C for 24 h in UHV. We interpret the abrupt decrease of the ZBC above 450°C as a consequence of the oxygen diffusion from Nb oxides into the bulk, leading to a cleaner interface and to a reduced inelastic scattering at the surface.

In the present study, we observe a shift of the quasiparticle peaks to voltages above the bulk Nb energy gap of 1.55 meV, suggesting an inelastic scattering mechanism. However, while baking the situation is more complex and well-known deleterious effects (massive injection of oxygen into Nb) do occur as well as other elastic processes that could lead to a filling of the gap. To extract the proportion of inelastic and elastic scattering mechanisms is beyond the scope of this paper and perhaps impossible. As a consequence, we think of $\Gamma$ as a broadening parameter that gives a quantitative estimate of the deviation of the measured superconducting DOS from an ideal one. The evolution of $\Delta$ and $\Gamma$ as a function of the baking temperature is summarized in Fig. 2a.

It is worth noting that for an unbaked coated sample, the fitting parameters are the same as for uncoated Nb samples, i.e., the deposition of an alumina overlayer by ALD doesn't modify the superconducting parameters at the surface of Nb. However, for baking temperatures of 250°C-2 h and 380°C-24 h, the superconducting DOS are strongly impaired, exhibiting reduced gap values of 1.35 and 1.3 meV respectively. The evolution of the gap coincides with the increasing peak intensity of the total niobium oxides extracted from the XPS fits (11) (Fig.2b), mostly composed of metallic NbO as the $Nb_2O_5$ is reduced. This result is in agreement with previous XPS studies (12) that indeed revealed a thickening of metallic NbO at the interface with the niobium. The smaller measured energy gap may thus be a result of a proximity effect between the Nb and a thicker NbO. The corresponding variations of the inelastic scattering value $\Gamma$, jumping from 0.6 meV at 30°C to 1 meV at 250°C-2h and decreasing back to 0.48 meV at 380°C-24h doesn't seem to correlate with the oxide composition. This can be explained however by the competition between the strong oxide dissociation above 200°C (13) and the diffusion of oxygen into Nb. At 250°C for 2 h, the diffusion length $L_{diff}$ calculated from the diffusion equation of oxygen into the bulk niobium, is 120 nm~$2\lambda_{Nb}$. Whereas above 380°C for longer time, 24 h, the diffusion is faster ($L_{diff} \geq 1mm$), and oxygen is driven far into the bulk. Consequently the $\Gamma$ value decreases. Interestingly, the evolution of the superconducting parameters correlates well with cavity performance after similar baking treatment (14).

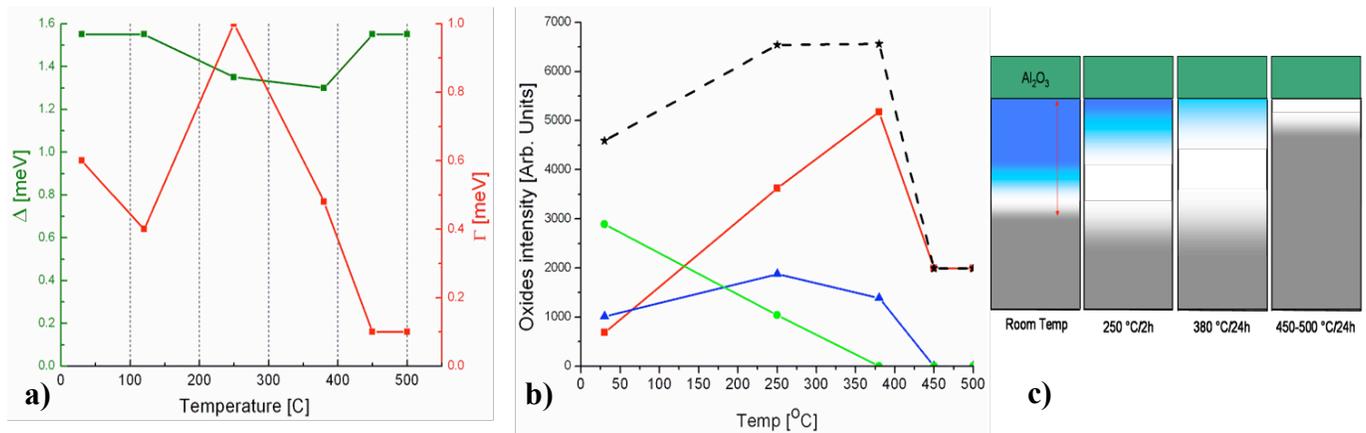

Fig.2: (Color online) **a)** evolution of the superconducting gap Δ and inelastic scattering parameter Γ as a function of the baking temperature. **b)** Plot of the intensity of Nb oxides peaks extracted from XPS fits, dashed line the total intensity of oxides peaks, in red the NbO peak intensity, blue the $NbO_2$ and in green the $Nb_2O_5$. **c)** Schematic evolution of the oxide composition at the surface of Nb.

Finally, for the highest baking temperature, >450°C for 24 h, the total intensity of niobium oxides peaks, composed of only NbO, decreases drastically in agreement with (12) and with the idea that oxygen diffuses far into the bulk. This result in a striking improvement of the superconducting properties: Γ=0.1 meV and Δ recovers the bulk Nb gap value of 1.55 meV. The saturation of the NbO peak intensity above 450°C may be an indication of a more stable and ordered NbO phase, as proposed by Hellwig (15), with a sharper NbO-Nb interface that would also contribute to improve the proximity effect of the superconducting DOS. Higher temperature anneals will be carried out on coated niobium coupons to search for further improvements in the superconducting DOS, aiming at surface properties of an ideal Nb superconductor with no inelastic scattering and optimum Nb cavity performance.

In conclusion, the overall evolution of the superconducting parameters, Γ and Δ, together with the oxides composition as a function of the baking temperature can be explained by a competition between Nb-oxide dissociation and oxygen diffusion into the bulk. Such a model has been suggested for low temperature (<150°C) anneals by G. Ciovati (13), but it has not been proven experimentally for high baking temperature. The close correlation of the results in Fig. 2a with cavity performance for the same temperature and baking time points to tunneling spectroscopy as a relevant tool to explain and predict cavity performance and inexpensively test new surface treatment protocols. A single cell 1.3 GHz cavity has been coated with 10 nm $Al_2O_3$ and 3 nm of $Nb_2O_5$ (to avoid multipacking effect) and RF performance test will be conducted soon.

This work was supported by US DOE, under contact DE-AC02-06CH11357.